\documentstyle[amsmath,epsf,epsfig,subfigure,preprint,aps,graphicx,amssymb,axodraw]{revtex}
\tighten \preprint{SNUTP 04-014}
\begin{document}
\title{\Large\bf A new mechanism for baryogenesis living
through electroweak era}
\author{
Hyung Do Kim$^{(a)}$\footnote{hdkim@phya.snu.ac.kr}, Jihn E.
Kim$^{(a)}$\footnote{jekim@phyp.snu.ac.kr}  and Takuya
Morozumi$^{(b)}$\footnote{morozumi@hiroshima-u.ac.jp} }
\address{
$^{(a)}$School of Physics, Seoul National University, Seoul
151-747, Korea\\
$^{(b)}$Graduate School of Science, Hiroshima University, Higashi
Hiroshima--739--8526, Japan} \maketitle

\begin{abstract}
We present a new mechanism for baryogenesis by introducing a heavy
vector-like SU(2)$_W$ singlet quark(s) with $Q_{\rm em}=\frac23$
quark $U$ or $Q_{\rm em}=-\frac13$ quark $D$. The lifetime of the
heavy quark is assumed to be in the range $2\times 10^{-11}\ {\rm
s}<\tau<1$ s. Being SU(2)$_W$ singlet, it survives the electroweak
phase transition era. It mixes with SU(2)$_W$ doublet quarks with
tiny mixing angles to satisfy the FCNC constraints, where a simple
$Z_2$ symmetry is suggested for realizing this scheme. The heavy
quark asymmetry is generated in analogy with the old GUT scenario.
\\
\vskip 0.5cm\noindent [Key words: Baryogenesis, FCNC, Sphaleron]
\end{abstract}


\newpage

\newcommand{\dis}[1]{\begin{equation}\begin{split}#1\end{split}\end{equation}}
\newcommand{\beqa}[1]{\begin{eqnarray}#1\end{eqnarray}}

\def\be{\begin{equation}}
\def\ee{\end{equation}}
\def\ben{\begin{enumerate}}
\def\een{\end{enumerate}}
\def\lsl{ l \hspace{-0.45 em}/}
\def\ksl{ k \hspace{-0.45 em}/}
\def\qsl{ q \hspace{-0.45 em}/}
\def\psl{ p \hspace{-0.45 em}/}
\def\ppsl{ p' \hspace{-0.70 em}/}
\def\dsl{ \partial \hspace{-0.45 em}/}
\def\Dsl{ D \hspace{-0.55 em}/}
\def\matrix{ \left(\begin{array} \end{array} \right) }

\def\ma{m_A}
\def\mf{m_f}
\def\mz{m_Z}
\def\mw{m_W}
\def\ml{m_l}
\def\ms{m_S}
\def\dag{\dagger}

\def\hf{\textstyle{\frac12~}}
\def\hff{\textstyle{\frac13~}}
\def\hfg{\textstyle{\frac23~}}
\def\DQ{$\Delta Q$}

\baselineskip 0.7cm
\bibliographystyle{plain}

\thispagestyle{empty}

The baryon asymmetry of universe is one of the most important
cosmological observations needed to be explained by a particle
physics model(s). Starting from a baryon symmetric universe,
Sakharov proposed three conditions for generating a baryon
asymmetry in the universe from fundamental interactions applicable
in cosmology \cite{sakharov}: the existence of baryon number
violating interactions which accompany C and CP violation, and
their working in a non-equilibrium state in the cosmos. With the
advent of grand unified theories(GUTs), some GUTs applied in the
evolution of the universe satisfied these three conditions
\cite{yoshimura}. A popular scenario was to have superheavy
colored scalars(${\bf 5}_H$s in SU(5) for example) decaying  at
least to two channels, say $qq$ and $ql$, both of which violate
the baryon number. Until mid-1980s, this GUT scenario seemed to be
the theory for the baryon asymmetry in the universe.

However, this GUT scenario underwent a nontrivial modification
after considering high temperature effects in nonabelian gauge
theories. The standard model(SM) has a non-perturbative sphaleron
solutions \cite{manton} whose effect on tunneling is supposed to
be extremely small, $\sim e^{-8\pi^2/g^2_W}$. But at high
temperature the sphaleron barrier for tunneling is overcome and it
was pointed out that the tunneling through sphaleron can be
important in cosmology \cite{shapo}. Since the sphaleron violates
the baryon number, it was argued that the baryon asymmetry
produced in the earlier epoch is erased during the epoch of
electroweak phase transition. In the GUT scheme, some models can
keep an asymmetry \cite{cko}. One example is to use the $B-L$
conservation of SO(10). Using the $B-L$ conservation during the
electroweak phase transition, the leptogenesis generating the $B$
number from a nonvanishing $L$ number was proposed \cite{FY}.
  The most recent complete
phenomenological analysis on the leptogenesis constrains the
lightest singlet Majorana neutrino mass $M_1>10^{9}$ GeV
\cite{buch}, probably contradicting the upper bound of the
reheating temperature after inflation in supergravity models
\cite{gravitino}. Therefore, theoretically and also for a
practical implementation of the baryogenesis, it is desirable to
invent any new mechanism for baryogenesis.

In this paper, we devise a mechanism for evading the sphaleron
$\Delta B$ erasing scenario. The mechanism employs a vector-like
SU(2)$_W$ singlet heavy quark $Q$, i.e. $Q_L$ and $Q_R$, which can
be $Q_{\rm em}=\frac23$ quark $U$ \cite{KK} or $Q_{\rm
em}=-\frac13$ quark $D$ \cite{Moro} so that it decays to ordinary
quarks through electroweak interactions. The SU(2)$_W$ singlet
vector-like quark(s) necessarily introduces flavor changing neutral
currents \cite{KK,GW} which is harmful if the heavy quark mass is
below the weak scale. If the heavy quark is sufficiently heavy and
survives through the epoch of the electroweak phase transition,
then the heavy quark asymmetry generated by the Sakharov
conditions would remain unwashed by the sphaleron processes. This
is because the 't Hooft interaction for an SU(2)$_W$
sphaleron does not involve SU(2)$_W$ singlets. Being SU(2)$_W$
singlets, a heavy quark asymmetry would not be erased. The
asymmetry in heavy quarks may be erased by the mixing with
SU(2)$_W$ doublet quarks. But the mixing angle $\epsilon$ in our
scheme is so small that the washing out of a heavy quark asymmetry by the
sphaleron is completely negligible. We will discuss more on this
later. Our proposal depends on whether one can construct a
theoretically reasonable model with SU(2)$_W$ singlet heavy quarks
with the following properties:
\begin{itemize}
\item The Sakharov conditions for creating $\Delta Q\sim \Delta B$
 are available. For $\Delta B\ne 0$ processes, GUTs are used. Also
 the CP violation at the GUT scale is present, and the heavy
 colored scalar particles are present for them to decay to colored
 quarks, anti-quarks and/or leptons.
\item The lifetime of $Q$ is sufficiently long,
$\tau_Q>O(10^{-11}\ {\rm s})$ \cite{tcosmic},
 so that the $Q$ asymmetry survives
the chaotic electroweak phase transition.
\item The heavy quark model must not introduce too large  flavor
changing neutral currents.
\end{itemize}

For definiteness, let us introduce $Q_{\rm em}=-\frac13$ quarks
$D$s.

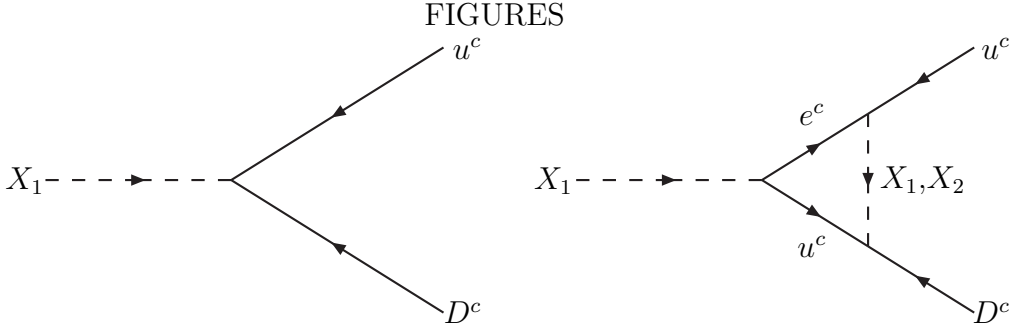
\begin{figure}[h]
\begin{center}
\begin{picture}(400,120)(0,-20)
\SetWidth{0.8}
 \DashArrowLine(30,50)(100,50){5}
\Text(15,50)[l]{$X_1$} \ArrowLine(180,100)(100,50)
\Text(195,100)[r]{$u^c$} \ArrowLine(180,0)(100,50)
\Text(195,0)[r]{$D^c$}

\DashArrowLine(230,50)(300,50){5} \Text(215,50)[l]{$X_1$}
\ArrowLine(300,50)(340,75) \Text(320,75)[]{$e^c$}
\ArrowLine(380,100)(340,75) \Text(395,100)[r]{$u^c$}
\ArrowLine(300,50)(340,25) \Text(320,25)[]{$u^c$}
\ArrowLine(380,0)(340,25) \Text(395,0)[r]{$D^c$}
\DashArrowLine(340,75)(340,25){5} \Text(378,50)[r]{$X_1$,$X_2$}
\end{picture}
\caption{The interference between tree and one-loop diagrams is
needed for a nonzero $\Delta Q$ generation. 
}\label{Fig}
\end{center}
\end{figure}

The first point is easily implementable in GUTs with SU(2)$_W$
singlet quarks such as in an E$_6$ GUT. In the E$_6$ GUT, one
family is embedded in ${\bf 27}_F$ which can contain 15 chiral
fields of the SM, a vector-like lepton doublets
$\{L_1(Y=-\frac12),L_2(Y=\frac12) \}$, one vector-like heavy quark
$D(Y=-\frac13)$ and $D^c(Y=\frac13)$, and two heavy neutrinos.
For three families, we introduce three ${\bf 27}_F$s. For the
Higgs mechanism, we introduce a scalar ${\bf 27}_H$ which contains
three Higgs doublets. An adjoint representation ${\bf 78}_H$ is
needed for breaking E$_6$ down to the SM. Certainly, there exists
a gauge hierarchy problem of how we remove most scalars of ${\bf
27}_H$ at the GUT scale, which is not addressed here. In this
setup, we have all the ingredients for the GUT baryogenesis
\cite{yoshimura}. Below, however, we will not restrict to the
E$_6$ GUT, but proceed to discuss in the SM framework with a heavy
$Q_{\rm em}=\frac13$ colored scalar $X_i$ ({ \rm anti-fundamental}
of $SU(3)_C$) to generate the heavy quark D number asymmetry \DQ\
with $i=1,2$. Relevant interactions are
\begin{eqnarray}
g_{Di} X_i u^c D^c + g_{ei} X_i^* u^c e^c + {\rm h.c.} .
\end{eqnarray}
All the fermions are written in terms of left-handed Weyl spinor.
($\psi^c$ is the charge conjugation of the right-handed spinor.)
The diagrams responsible for the \DQ\ asymmetry in the decay of
heavy colored scalar fields are shown in Fig. \ref{Fig}.

The decay of $X_i$ ($X_i^*$) gives a positive (negative) \DQ\
number. We can consider all possible baryon number generations,
but those light-quark numbers are washed out during the
electroweak phase transition through the sphaleron process and
only \DQ\ number is the meaningful conserved quantity.

If we had just a single colored scalar $X$ ($i=1$), the first and
the second diagrams in Fig. \ref{Fig} are proportional to $g_D$
and $g_e^* g_e g_D$, respectively, and
the crossing term $g_e^* g_e g_D^* g_D$ is real. In
this case we should consider higher loop corrections which are
highly suppressed \cite{segre}. The leading interference term can
be complex  if we have two colored scalars $X_i$ with $i=1,2$. The
phase from this leading interference term is proportional to
\begin{eqnarray}
{\rm arg} \left( g_{D1}^* g_{D2} g_{e1}^* g_{e2}
\right).\label{phase}
\end{eqnarray}
If we allow arbitrary phases in the Yukawa couplings, the relative
phase of $g_{D1}$ and $g_{D2}$ can be canceled only by the
relative phase redefinition of $X_2$ compared to $X_1$. It is also
true for $g_{e1}$ and $g_{e2}$. Therefore, if ${\rm arg}
(g_{D1}/g_{D2})$ is different from ${\rm arg} (g_{e1}/g_{e2})$,
one of the phases in the Yukawa couplings cannot be rotated away.
This proves that the phase appearing in the interference
(\ref{phase}) is physical (i.e. {\it un-removable}).

In principle, we can consider decays of both $X_1$ and $X_2$.
However, we will assume that $X_2$ is heavier than $X_1$ and the
decay of $X_1$ dominates.

$\Delta Q$ number generated from the above decay is, including the
contribution from the wave function correction diagram 
\cite{Liu:1993tg},
\begin{eqnarray}
\frac{n_D}{s} & \simeq & \frac{\kappa}{4\pi g_*} 
\frac{{\rm Im} ( g_{D1}^* g_{D2} g_{e1}^* g_{e2} 
)}{(g_{D1}^* g_{D1} + g_{e1}^* g_{e1})} 
\left[ f_v (x) + f_s (x) \right],
\end{eqnarray}
where $\kappa$ is the washout factor, $f_v(x) = 1 
- x \log (1+\frac{1}{x})$,
$f_s(x) = \frac{1}{x-1}$, $x=\frac{m_{X_2}^2}{m_{X_1}^2}$ 
is the mass ratio of two heavy scalar fields and $g_* = 106.75$ is 
the number of effective degrees of freedom in the SM. 
Within the range of parameters we discuss, it is possible to obtain 
the observed value $n_B/s = (8.7 \pm 0.4 )\times 10^{-11}$ in our 
scheme. Generation of $\Delta Q$ number after inflation will 
be discussed later.

For a very long lifetime $\tau_D$, the mixing angle between the
ordinary quarks and heavy quarks $D$ must be sufficiently small.
The estimation of the mixing angle dependence on the quark masses
can be considered in the following way. Introducing a discrete
symmetry, e.g. by giving different discrete quantum numbers to
$b_R$ and $D_R$, one can consider the following $2\times 2$
simplified mass matrix for an ordinary quark $b$ and a heavy
singlet quark $D$,
\begin{equation}
A=\left(
\begin{array}{cc}
m & J\\
0 & M
\end{array}\right)\label{twobytwo}
\end{equation}
where $J$ parametrizes the $D_R$ coupling to $b_L$. The
eigenvalues of $A A^{\dagger}$
are found to be
\begin{eqnarray}
\left(
\begin{array}{c}
 |m_b|^2\\ |m_D|^2\end{array} \right)=\hf ({|M|^2+|m|^2+|J|^2})
\mp \hf{\sqrt{[(|M|+|m|)^2+|J|^2][(|M|-|m|)^2+|J|^2]}}
\end{eqnarray}
which, in the limit $|M|^2\gg |m|^2,|mJ|$, approximate to
\begin{eqnarray}
|m_b|\simeq |m| ,\ \ |m_D|\simeq |M|
\end{eqnarray}

The Hermitian matrix $AA^\dagger$ is diagonalized by a unitary
matrix. Taking  vanishing phases, in the super-large $M$ limit we
obtain the eigenstates,
\begin{equation}
|b\rangle\simeq \left(\begin{array}{c}1\\ -\frac{J}{M}
\end{array}\right),\ \ |D\rangle\simeq \left(\begin{array}{c}
\frac{J}{M}\\ 1 \end{array}\right).\label{small1}
\end{equation}
From Eq. (\ref{small1}), one can make the mixing sufficiently
small by taking $J/M\rightarrow 0$.

For generalizing to three ordinary quarks $d_i\ (i=1,2,3)$ and $n$
heavy quarks $D_J\ (J=1,\cdots,n)$, one can consider the following
$(3+n)\times(3+n)$ mass matrix {\bf M} to
$$
{\bf M}=\left(
\begin{array}{cc}
M_d & J\\
 J^\prime & M_D
\end{array}
\right)
$$
where $M_d$ is a $3\times 3$ mass matrix for ordinary $Q_{\rm
em}=-\frac13$ down-type $d$ quarks , $J$ is a $3\times n$ matrix,
$J^\prime$ is an $n\times 3$ matrix, and $M_D$ is an $n\times n$
mass matrix for the heavy $Q_{\rm em}=-\frac13$ quarks $D$s. We
can redefine the right-handed fields only to make $J^{\prime}$
vanish. If $J^{\prime}/M_D$ is small and $M_d \sim J$, there is
no correction to the matrix. Thus without loss of generality, we
can consider the following mass matrix
\begin{equation}
{\bf M}=\left(
\begin{array}{cc}
M_d & J\\
 {\bf 0} & M_D
\end{array}
\right)\label{Mstand}
\end{equation}
where $n\times 3$ elements of matrix {\bf 0} are zeros. If the
elements of $M_d$ are O($m$), the elements of $J$ are O($J$), and
the elements of $M_D$ are O($M$), the mixing angles between $d$
and $D$ quarks are of order $\epsilon\sim J/M$. Thus, if
$\epsilon$ is sufficiently small then the lifetime(s) of $D$
quark(s) can be made long.

For an order of magnitude estimation for the lifetime of the
lightest $D$, let us use the $b-D$ system with negligible
couplings to the other light quarks, viz. Eq. (\ref{twobytwo}).
The flavor changing neutral current coupling of $Z_\mu$ is of
order $\epsilon$ also. Thus, we can estimate the decay width of
$D$ from $D\rightarrow tW, bZ, bH^0$ as
\begin{eqnarray}
\Gamma_D&=&\frac{  \sqrt{2} G_F} {8 \pi} J^2 m_D.
\end{eqnarray}
where we assumed $m_D\gg m_t$. $J$ is parametrized as
\begin{equation}
|J|\equiv fm_b
\end{equation}
 with $f$ a small number. Here, $\epsilon\equiv fm_b/m_D$.  
The lifetime of $D$ should be made longer than $2\times 10^{-11}$
s for $D$ to pass through the electroweak phase transition era.
However, it should not be too long, say $\tau_D<1$ s, not to
disrupt the standard nucleosynthesis. Therefore, we require the
following {\it cosmological lifetime window} for the lightest $D$,
\begin{equation}
2\times 10^{-11}\ {\rm sec}\le \tau_D \le 1\ {\rm
sec}\label{bound}
\end{equation}
The lifetime window given above can be translated into the
constraint equation on $|\epsilon|$
\begin{equation}\label{boundeps}
 \frac{1}{\left(10^6 \ m_D ({\rm GeV})
  \right)^{3/2}} \le |\epsilon|  \le
 \frac{1}{\left(2.7\times 10^{2} \ m_D ({\rm GeV}) \right)^{3/2}}.
 \end{equation}
For a large $m_D$, on the other hand, the four fermion interaction 
is the dominant one. Then, the condition (D decay rate) $\le H$ at
$T\sim M_W$ is $m_D\le 10^5(M_{X_1}/10^{10}\ {\rm GeV})^{4/5}$ GeV.

To estimate how much of $\Delta D$ is washed out by the sphaleron
processes, we first estimate the mixing of mass eigenstates $b_L$
and $D_L$ in the weak eigenstate $b_L^0$: $b_L^0\simeq
b_L+\epsilon D_L$.  The oscillation
period  is $\sim 1/m_D$. The probability
to find SU(2)$_W$ doublet $b_L^0$ from $D_L$ is $\sim
|\epsilon|^2$ in the time interval $1/m_D$, or the rate to go to
$b_L^0$ is $m_D|\epsilon|^2$. Since the period of the electroweak
phase transition lasts for $1/H \sim M_P/M_W^2$, 
the amount of $\Delta Q$ washed out during the electroweak phase 
transition is $(m_D M_P/M_W^2)|\epsilon|^2$. This condition gives
a rough bound, $|\epsilon|\le 10^{-8}$ for $m_D\simeq (100-1000)$ GeV.
For $m_D>$ 1 TeV, Eq. (\ref{boundeps}) gives a stronger bound.

The model presented above is constrained by the FCNC and proton
decay experiments. The mass of $D$ is constrained  by the bounds
on FCNC.

The mixing of $D$ quark with $b$ quark may change the flavor diagonal
coupling $Z \to b \bar{b}$ from the standard model one
\begin{equation}
z_{bb}=1-|\epsilon|^2.
\end{equation}
The experimental bound on $z_{bb}=0.996 \pm 0.005$ leads to
$|\epsilon|^2 \le 0.009$. \cite{Aguila}
If $D$ quark mixes with $b$ and $s$ with the strength $J_b$ and $J_s$,
rare B decay $ B \to Xs l^+ l^- $ occurs in the tree level.
The bound obtained from the analysis \cite{mxy}
\begin{equation}
|z_{sb}|=\frac{J_b J_s}{m_D^2}< 1.4 \times 10^{-3}.
\end{equation}
Finally, we consider FCNC constrains obtained from Kaon system. We
require the FCNC contribution is smaller than the standard model.
From the $K_L$ and $K_S$ mass difference mass difference,
\begin{equation}
|z_{sd}| \le \left(\frac{G_F m_c^2 \lambda^2}{2 \sqrt{2} \pi^2}\right)^{1/2}.
\end{equation}
From the $K^+ \to \pi^+ \nu {\bar \nu}$
\begin{equation}
\frac{Br(K^+ \to \pi^+ \nu {\bar \nu})|_{FCNC}}{Br(K^+ \to \pi^0 e^+ \nu)}
=\frac{3}{2} \frac{|z_{sd}|^2}{\lambda^2}
 \le 2 \times 10^{-9}
\end{equation}
This leads to
\begin{equation}
|z_{sd}| \le 7.3 \times 10^{-6}.
\end{equation}
If we neglect the flavor dependence of $J$, the upper bounds on
$\epsilon$ obtained from $ Z \to b \bar{b}$, $B \to Xs l^+ l^-$,
$\Delta m_K$ and $ K^+ \to \pi^+ \nu \bar{\nu}$ are $0.037$,
$0.095$, $0.02$ and $7.3 \times 10^{-6}$, respectively. The
tightest constraint from $ K^+ \to \pi^+ \nu \bar{\nu}$, gives
$m_D \ge 6.6 \times 10^{6} f$  GeV from $|\epsilon|\simeq
fm_b/m_D$. Then, Eq. (\ref{boundeps}) gives
\begin{equation}
\frac{1}{4.8\times 10^9 \sqrt{m_D(\rm
GeV)}}<|f|<\frac{1}{2.1\times 10^4 \sqrt{m_D(\rm GeV)}}
\end{equation}
which can be satisfied by some range of small couplings. The $J$
term is supposed to arise from breaking a $Z_2$ symmetry, and its
smallness can be implemented naturally.

The highly suppressed $J$ can be obtained by
introducing a discrete symmetry. Consider a $Z_2$ symmetry under
which the quarks have the following charges (D parity),
\begin{equation}
Z_2:\ b_{L,R}\rightarrow b_{L,R}, \
 D_{L,R} \rightarrow -D_{L,R}.
\end{equation}
This implies that the SU(2)$_W$ doublet $q_L$ housing $b_L$ has
the same $Z_2=+1$ eigenvalue. The SU(2)$_W$ singlet $D$ can have a
bare mass or obtain a mass by a large VEV of $Z_2=+1$ singlet
scalar $S$. All the interactions are consistent with D parity if
$L, e^c$ and $X$ are D parity odd and all the other fields are
even. We introduce two Higgs doublets, $\phi$ and $\varphi$,
\begin{equation}
Z_2:\ \phi\rightarrow \phi,\ \varphi\rightarrow -\varphi.
\end{equation}
If this $Z_2$ symmetry were exact, we obtain $J=0$. Here we
introduce a small amount of $Z_2$ breaking by a soft term
$m_\delta^2$,
\begin{eqnarray}
V(\phi,\varphi)&=&(m_\delta^2\varphi^\dagger\phi+h.c.)
-\mu^2\phi^\dagger\phi +M^2_{\varphi}\varphi^\dagger\varphi
+\lambda_1(\varphi^\dagger\varphi)^2+\lambda_2(\phi^\dagger\phi)^2
\nonumber\\
&&+\lambda_3\varphi^\dagger\varphi\phi^\dagger\phi
+(\lambda_4\varphi^\dagger\phi\phi^\dagger\varphi
+\lambda_5\varphi^\dagger\phi\varphi^\dagger\phi+h.c.)
\end{eqnarray}
where we assume $M_\varphi^2\gg\mu^2\gg|m_\delta^2|>0.$ Then,
$\phi$ develops a VEV $v$ for the electroweak symmetry breaking,
but $\varphi$ does not if $m^2_\delta=0$. For a nonzero
$m^2_\delta$, $\varphi$ develops a tiny VEV,
$\langle\varphi\rangle\simeq vm_\delta^2 /M^2_\varphi$. Thus, the
coupling $f_{\rm off}\bar q_L\varphi D_R$ gives $J\sim f_{\rm
off}vm_\delta^2 /M^2_\varphi\sim (\sqrt2 f_{\rm
off}/f_b)m_bm_\delta^2 /M^2_\varphi$. Then, our small parameter
$f$ is $f=(\sqrt2 f_{\rm off}/f_b)m_\delta^2 /M^2_\varphi$. A
small soft mass can lead to a very small $f$ even for O(1) value
of $f_{\rm off}$. One loop diagram can generate $J$ which must be
proportional to $m^2_\delta$; hence it is subdominant. By the
choice of the soft $Z_2$ breaking term, $J$ can be made to fall in
the region for the needed lifetime window (\ref{bound}) for $D$.

Assuming no approximate discrete symmetry, proton decay can
proceed via the $X$ particle exchange $d^cu^c\rightarrow X^*
\rightarrow ue$. Thus, the mass of $X$ particle should be in the
GUT scale with a small Yukawa couplings to the first family
members as studied in GUT proton decay. This is the standard
colored Higgs mediated proton decay. But the mass of $D$ is not
restricted by proton decay.

But in the inflationary scenario, the $X$ particle should be light
enough($<10^{13}$ GeV) so that enough $X$s are present after
inflation. In supersymmetric models, one may need a stronger
constraint, $M_X<10^{9-10}$ GeV, from the gravitino problem
\cite{gravitino}. To implement this constraint, one can use the
above softly broken $Z_2$ symmetry with $X$ carrying $Z_2=-1$
parity. Thus, the proton decay operator $ud\rightarrow X
\rightarrow u^ce^+$ has an additional suppression factor $\xi^2$
which is nonvanishing only if the $Z_2$ is broken. For
$\tau_p>10^{33}$ years, we obtain $\xi<0.7\times 10^{-6}$ for
$M_X=10^{10}$ GeV. Preheating scenario \cite{preheating} can generate
sizable baryon asymmetry even for heavy $X$ (heavier than inflaton).
It can happen if $\Gamma_X \le 10^{-3} M_X$ which can be satisfied for
$g_{1(e,D)} \sim 10^{-1}$. Therefore, we can generate sizable
baryon asymmetry
from $X$ decay after inflation.

In conclusion, we devised a new mechanism for baryogenesis by
introducing SU(2)$_W$ singlet quark(s). The conditions for the
current mechanism to work are to generate $\Delta Q$ at the GUT
scale and the lifetime  bound must fall in the region
(\ref{bound}).

\begin{acknowledgments}
This work is supported in part by the KOSEF Sundo Grant(J.E.K.),
the ABRL Grant No. R14-2003-012-01001-0(H.D.K., J.E.K.),  the BK21
program of Ministry of Education, Korea(H.D.K., J.E.K.), and also
by the kakenhi of MEXT, Japan, No. 13640290(K.M.). J.E.K. and K.M.
also thank the Yukawa Institute for Theoretical Physics at Kyoto
University, where this work was initiated during the workshop
YITP-W-04-08 on $\lq\lq$Summer Institute 2004".
\end{acknowledgments}


\begin{thebibliography}{99}
\bibitem{tcosmic} The cosmic time $t=2.42\times 10^{-6}\ {\rm s}
/\sqrt{N}T_{\rm GeV}^2$ at the electroweak phase transition
$100-300$ GeV  is in the range $2\times (10^{-11}-10^{-12})$
second with $N= $O(100).
\end{thebibliography}
\end{document}